\documentclass[usenatbib, usegraphicx, useAMS]{mn2e}

\usepackage{times}

\title[Hamiltonian Power Spectrum Estimation]{Fast optimal CMB power spectrum
  estimation with Hamiltonian sampling} \author[Taylor, Ashdown \&
Hobson]{J.~F.~Taylor \thanks{E-mail:j.taylor@mrao.cam.ac.uk},
  M.~A.~J.~Ashdown and M.~P.~Hobson \\ Astrophysics Group, Cavendish
  Laboratory, JJ Thomson Avenue, Cambridge CB3 0HE, UK}

\date{Accepted ---. Received ---; in original form \today}
\pagerange{\pageref{firstpage}--\pageref{lastpage}}
\pubyear{2006}

\newcommand{\pr}[1]{\rmn{Pr}\left( #1 \right)}
\newcommand{\mat}[1]{\mathbfss{#1}}

\begin{document}
\label{firstpage}
\maketitle
\begin{abstract}
  We present a method for fast optimal estimation of the temperature
  angular power spectrum from observations of the cosmic microwave
  background.  We employ a Hamiltonian Monte Carlo (HMC) sampler to
  obtain samples from the posterior probability distribution of all
  the power spectrum coefficients given a set of observations. We
  compare the properties of the HMC and the related Gibbs sampling
  approach on low-resolution simulations and find that the HMC method
  performs favourably even in the regime of relatively low
  signal-to-noise. We also demonstrate the method on high-resolution
  data by applying it to simulated \textit{WMAP} data.  Analysis of a
  \textit{WMAP}-sized data set is possible in a around eighty hours on
  a high-end desktop computer. HMC imposes few conditions on the
  distribution to be sampled and provides us with an extremely
  flexible approach upon which to build.
\end{abstract}
\section{introduction}
\label{sec:intro}
Observations of the cosmic microwave background (CMB) have proved to
be extremely valuable for testing and constraining cosmological
models. The majority of models predict that the anisotropies in the
CMB signal are Gaussian and their statistics isotropic across the
sky. The angular power spectrum \(C_{\ell}\) therefore provides a
natural connection between theory and observation and a variety of
methods have been explored to compute the power spectrum from sets of
observations.

Maximum-likelihood methods \citep*{Gorski1994a, Bond1998, Oh1999}
provide an optimal estimate of the CMB power spectrum which has made
them an invaluable tool for analysing the CMB for single-dish
experiments and interferometers \citep{Hobson2002}. Brute force
implementations of the method can only be applied to small data sets
as the required computation scales as
\(\mathcal{O}(N_{\rmn{pix}}^{3})\), where \(N_{\rmn{pix}}\) is the
number of pixels in a CMB map (see \citet{Efstathiou2003} for a
review). For a number of special cases one can construct
maximum-likelihood estimators that perform more favourably
\citep{Challinor2002, Wandelt2003}, although their lack of generality
limits their applicability and, as their computational demands scale
as \(\mathcal{O}(N_{\rmn{pix}}^{2})\), even they cannot be applied
directly the largest contemporary (\textit{WMAP}) or future
(\textit{Planck}) data.

Alternatively one can resort to approximate pseudo-\(C_{\ell}\)
methods, \citet{Hivon2002}. These scale as the map-making process and
are fast even for the largest data sets. Hybrid methods
\citep{Efstathiou2004} combine a maximum-likelihood approach on large
angular scales with a fast pseudo-\(C_\ell\) estimator on small
scales.

To compare theoretically predicted power spectra and those estimated
from a set of observations it is necessary to construct a likelihood
function. Maximum-likelihood and pseudo-\(C_{\ell}\) methods can only
provide approximations to this likelihood.

An alternative framework has been developed \citep*{Wandelt2004,
  Jewell2004} where one explores the full posterior distribution of
the power spectrum with Monte Carlo samples. This method is not only
exact but scales like the pseudo-\(C_\ell\) methods. Under the
assumption of position invariant, circularly symmetric beams and
uncorrelated noise, one can perform the beam convolution in the
spherical harmonic domain and evaluate the likelihood of the data in
the map domain, and the method scales as
\(\mathcal{O}(N_{\rmn{pix}}^{3/2})\). The favourable scaling has
enabled the method to be applied to the \textit{WMAP} data
\citep{Bennett2003, Eriksen2004}.

The approach relies on the availability of an efficient method for
sampling from high-dimensional distributions. Previous implementations
use a Gibbs sampler but this restricts the applicability of the method
to Gaussian noise and CMB. We propose the use of a Hamiltonian Monte
Carlo (HMC) sampler \citep{Duane1987}. As opposed to the majority of
Markov-Chain Monte Carlo (MCMC) methods, HMC scales well with problem
size.  Few requirements are made on the distribution to be sampled,
thus giving us the opportunity for great flexibility. HMC has been
widely applied in Bayesian computation \citep{Neal1993} and has also
been employed for cosmological parameter estimation
\citep{Hajian2007}.

In this work we begin, in Section \ref{sec:psest}, by outlining the
procedure for estimating power spectra with sampling. In Section
\ref{sec:hmc} we describe the HMC method and a technique for
determining the convergence of samples drawn with a HMC sampler.  A
summary of the process of applying HMC to the power spectrum
estimation problem can be found in Section \ref{sec:hmcpsest} and we
provide a prescription for setting the many tuneable parameters of the
sampler. In Section \ref{sec:lowres} we apply the method to
low-resolution simulations and compare the Hamiltonian and Gibbs
samplers. Section \ref{sec:wmap} details our application of the method
to simulated \textit{WMAP} observations. Our conclusions are presented
in Section \ref{sec:conc}.
\section{power spectrum estimation with sampling}
\label{sec:psest}
Suppose the true CMB sky, divided for convenience into pixels, is
represented by the temperature vector \(\bmath{t}\).  The sky is
observed and the resultant data vector \(\bmath{d}\), in any domain,
is the sum \(\bmath{d} = \bmath{s} + \bmath{n}\) of contributions due
to the underlying CMB signal \(\bmath{s}\) in that domain and the
corresponding noise \(\bmath{n}\). Moreover the signal \(\bmath{s}\)
is usually linearly related to the true CMB sky \(\bmath{t}\). Thus we
have
\begin{equation}
  \label{eq:skymodel}
  \bmath{d}=\mat{R}\bmath{t}+\bmath{n},
\end{equation}
where the matrix \(\mat{R}\) represents the linear mapping from the
true CMB sky to the corresponding CMB signal in whatever domain the
data resides.

In the following discussion, we need not assume a particular domain
for the generic data vector \(\bmath{d}\). Nevertheless, it is most
common for \(\bmath{d}\) to represent the pixelised CMB map convolved
with the instrument beam and our work, so far, has used solely this
form for the data vector.

The temperature field \(\bmath{t}\) is related to the spherical
harmonic coefficients of the field \(\bmath{a}\) by
\begin{equation}
  \label{eq:sht}
  t\left(x_p\right) = 
    \sum_{\ell=2}^{\ell_{\rmn{max}}}
    \sum_{m=-\ell}^{\ell}a_{\ell m}Y_{\ell m}\left(x_p\right),
\end{equation}
where \(t\left(x_p\right)\) is a single pixel in the map vector
\(\bmath{t}\) and the \(Y_{\ell m}\) are the spherical harmonics.
Although formally one may take the upper limit of the \(\ell\)
summation to be infinite, it is more typical to choose a finite value
for \(\ell_\mathrm{max}\) appropriate to the beam size. We have not
considered the effect of the mono- and dipole contributions, the
handling of which, within this framework, is discussed in
\citet{Eriksen2004}. In this notation we may write our model for the
data in the form
\begin{equation}
  \label{eq:skymodelalm}
  \bmath{d}=\mat{R}\mat{Y}\bmath{a}+\bmath{n},
\end{equation}
where \(\mat{Y}\) describes the application of the spherical harmonic
transform and we represent the spherical harmonic coefficients by a
real vector.

For an isotropic Gaussian CMB sky the covariance matrix \(\mat{C}\) of
the \(a_{\ell m}\) has components
\begin{equation}
  \label{eq:pstoalms}
  C_{\ell m \ell^{\prime} m^{\prime}} =
  \langle a_{\ell m} a_{\ell^\prime m^\prime}^{*}\rangle =
  C_\ell \delta_{\ell \ell^\prime} \delta_{m m^\prime},
\end{equation}
where the set of coefficients \(\{C_{\ell}\}\) constitute the
theoretical angular power spectrum. Note that, since the sky is real,
\(a_{\ell m} = a_{\ell, -m}^{*}\)

We aim to sample from the joint distribution of the power spectrum
coefficients \(\pr{\{C_\ell\} | \bmath{d}}\). Although this is
difficult to perform directly, it is possible to sample from the joint
density of the power spectrum coefficients and the signal realization
\( \pr{\{C_\ell\} , \bmath{a}|\bmath{d}}\) and then marginalise over
\(\bmath{a}\). The joint density can be written as the product of the
appropriate conditional distributions
\begin{equation}
  \label{eq:joint}
  \pr{\{C_\ell\}, \bmath{a}|\bmath{d}} \propto 
  \pr{\bmath{d}|\bmath{a}}
  \pr{\bmath{a}|\{C_\ell\}}
  \pr{\{C_\ell\}}.
\end{equation}
The choice of prior \(\pr{\{C_\ell\}}\) is an interesting topic.
\citet{Wandelt2004} have some suggestions for making this choice but
for the purpose of this work we set \(\pr{\{C_\ell\}} = 1\) so that the
maximum of our posterior will correspond directly to a
maximum-likelihood estimate.

Given our choice of prior and assuming the noise is Gaussian then the
conditional distributions that make up (\ref{eq:joint}) can be written
in the form
\begin{equation}
  \label{eq:pagivend}
  \pr{\bmath{d}|\bmath{a}} \propto 
  \exp\left[-\frac{1}{2}
    \left(\bmath{d}-\mat{R}\mat{Y}\bmath{a}\right)^{\rmn{T}}
    \mat{N}^{-1}
    \left(\bmath{d}-\mat{R}\mat{Y}\bmath{a}\right)
  \right],
\end{equation}
where \(\mat{N} = \langle \bmath{n} \bmath{n}^{\rmn{T}} \rangle\), and 
\begin{equation}
  \label{eq:pagivencl}
  \pr{\bmath{a}|\{C_\ell\}} \propto
  \frac{1}{\sqrt{|\mat{C}|}}
  \exp
  \left(
    -\frac{1}{2} \bmath{a}^{\rmn{T}}\mat{C}^{-1} \bmath{a}
  \right) 
\end{equation}
where \(\mat{C}\) is easily constructed using (\ref{eq:pstoalms}). It
is convenient to rewrite this in the form
\begin{equation}
\label{eq:pagivenclalt}
\pr{\bmath{a}|\{C_\ell\}} \propto 
\prod_{l=2}^{\ell_{\rmn{max}}} \left(\frac{1}{C_{\ell}}\right)^{\frac{2\ell+1}{2}}
\exp \left(-\frac{2\ell+1}{2}\frac{\sigma_{\ell}}{C_{\ell}}\right),
\end{equation}
where \(\sigma_\ell = \frac{1}{2\ell + 1} \sum_m |a_{\ell m}|^2\) is
the power spectrum of the signal realization.

The selection of a domain in which to represent the data is determined
by the requirement that \(\mat{N}\) has a simple form. In this work we
make the assumption that in the map domain \(\mat{N}\) is well
represented by a diagonal matrix. In this domain incomplete sky
coverage is straightforwardly handled by setting the elements of
\(\mat{N}^{-1}\) that correspond to excluded pixels to zero. If the
instrument beam is position invariant and circularly symmetric then we
can compute the beam convolution quickly in harmonic space and the
predicted noiseless data can be written in the form \( \mat{Y}\mat{B}
\bmath{a}\) where \(\mat{B}\) represents the smoothing by the beam.

The computational cost of evaluating the posterior (and its gradients)
is now limited by the speed at which one can compute the spherical
harmonic transform \(\mat{Y}\). The transforms scale as
\(\mathcal{O}(N_{\rmn{pix}}^{3/2})\) and can be efficiently
parallelised.

We draw samples from the joint space \((\bmath{a},\{C_\ell\})\) using a
Hamiltonian Monte--Carlo sampler described in Section \ref{sec:hmc}.
\section{Hamiltonian Monte Carlo}
\label{sec:hmc}
Let us suppose that we wish to draw samples from a target density
\(\pr{\bmath{x}}\), where \(\bmath{x}\) is the \(N\)-dimensional
vector of our parameters. Conventional MCMC methods move through the
parameter space by a random walk and therefore require a prohibitive
number of samples to explore-high dimensional spaces. The Hamiltonian
Monte Carlo method \citep{Duane1987, Neal1993, Neal1996} draws
parallels between sampling and classical dynamics. By exploiting
techniques developed for describing the motion of particles in
potentials it is possible to suppress random walk
behaviour. Introducing persistent motion of the chain through the
parameter space allows HMC to maintain a reasonable efficiency even
for high dimensional problems \citep{Hanson2001}.

For each parameter, \(x_i\) we introduce a `momentum' \(p_i\) and a
`mass' \(m_i\); we discuss how to set the mass in the Appendix. We
construct a Hamiltonian formed from a potential energy term \(\psi
\left(\bmath{x}\right)\) and a kinetic energy term such that
\begin{equation}
  \label{eq:ham}
  H = \sum_i \frac{p_i^2}{2m_i} + \psi \left(\bmath{x}\right),
\end{equation}
where our potential is related to the target density by
\begin{equation}
  \label{eq:psi}
  \psi \left(\bmath{x}\right) = - \log \pr{\bmath{x}}.
\end{equation}
Our new objective is to draw samples from a distribution that is
proportional to \(\exp \left( -H\right)\). The form of the Hamiltonian
is such that this distribution is separable into a Gaussian in
\(\bmath{p}\) and the target distribution, i.e.
\begin{equation}
  \label{eq:hmchamsep}
  \exp \left( -H\right) =  \pr{\bmath{x}}
  \prod_i \exp \left( - \frac{p_i^2}{2m_i} \right).
\end{equation}
We can then obtain samples from \(\pr{\bmath{x}}\) by marginalising
over \(\bmath{p}\).

To find a new sample we first draw a set of momenta from the
distribution defined by our kinetic energy term, i.e.  an \(N\)
dimensional uncorrelated Gaussian with a variance in dimension \(i\)
of \(m_i\). We then allow our system to evolve deterministically, from
our starting point \(\left(\bmath{x}, \bmath{p}\right)\) in the phase
space for some fixed time \(\tau\) according to Hamilton's equations,
\begin{eqnarray}
  \label{eq:hamiltons}
  \frac{\rmn{d} x_i}{\rmn{d} t}& =& \frac{\upartial H}{\upartial p_i}\\
  \frac{\rmn{d} p_i}{\rmn{d} t} &= &- \frac{\upartial H}{\upartial x_i} =
  -\frac{\upartial \psi \left(\bmath{x}\right)}{\upartial x_i}.
\end{eqnarray}
At the end of this trajectory we have reached the point
\(\left(\bmath{x}^\prime, \bmath{p}^\prime \right)\) and we accept
this point with probability
\begin{equation}
  \label{eq:acceptance}
  p_A = \min \left( 1, \exp \left( -\delta H \right) \right),
\end{equation}
where
\begin{equation}
  \label{eq:dH}
  \delta H = H\left(\bmath{x}^\prime, \bmath{p}^\prime\right) - 
   H\left(\bmath{x}, \bmath{p}\right).
\end{equation}
After a new proposed sample is generated the momentum variable is
discarded and the process restarts by randomly drawing a new set of
momenta as described above.

This implies that if we are able to integrate
Hamilton's equations exactly then, as energy is conserved along such a
trajectory, the probability of acceptance is unity.

In fact the method is more general as, provided one uses the
Metropolis acceptance criterion (\ref{eq:acceptance}), it is permitted
to follow any trajectory to generate a new candidate point. However
only trajectories that approximately conserve the value of the
Hamiltonian (\ref{eq:ham}) will result in high acceptance rates. For
some problems it may be advantageous to generate trajectories using an
approximate Hamiltonian that can be computed rapidly, and bear the
cost of lowering the acceptance probability.

To integrate the equations of motions it is common practice to use the
leapfrog method. This method has the property of exact reversibility
which is required to ensure the chain satisfies detailed balance. It
is also numerically robust and allows for the simple propagation of
errors. We make \(n\) steps with a finite step size \(\epsilon\), such
that \(n\epsilon = \tau\), as follows,
\begin{eqnarray}
  \label{eq:leapfrog}
  p_i\left(t + \frac{\epsilon}{2}\right) &= &
  p_i\left(t\right) - 
  \frac{\epsilon}{2}
  \frac{\upartial \psi \left(\bmath{x}\right)}
  {\upartial x_i}\bigg{\vert}_{\bmath{x}\left(t\right)}\\
  x_i\left(t + \epsilon\right)&=&
  x_i\left(t\right) + 
  \frac{\epsilon}{m_i}p_i\left(t + \frac{\epsilon}{2}\right)\\
  p_i\left(t + \epsilon\right) &= &
  p_i\left(t + \frac{\epsilon}{2}\right) -  
\frac{\epsilon}{2}\frac{\upartial \psi\left(\bmath{x}\right)}
{\upartial x_i}\bigg{\vert}_{\bmath{x}\left(t+\epsilon\right)}
\end{eqnarray}
until \(t = \tau\). The interval \(\tau\) must be varied, usually by
drawing \(n\) and \(\epsilon\) randomly from uniform distributions, to
avoid resonant trajectories.  Higher-order integration schemes are
permitted, provided exact reversibility is maintained, although
generally incur significant additional computational costs.
\subsection{Convergence tests}
\label{sec:converge}
Diagnosing the convergence of a chain in an MCMC process is the
subject of much literature (see \citet{Cowles1996, Brooks1997} for
comprehensive reviews). \citet{Hanson2001} provides a method that uses
the gradient information, which we must possess to calculate
trajectories in HMC, to compute a convergence criteria.

One constructs two estimates of the variance of a chain, that depend
quite differently upon the distribution of samples across the target
density, although the basic method is easily generalised to
(combinations of) higher order central moments of
\(\pr{\bmath{x}}\). When the two estimates agree to within a certain
accuracy the chain is assumed to have converged.

We compute the variance of each parameter \(x_i\) independently. Our
first estimate of the variance of the samples is calculated by
\begin{equation}
  \label{eq:var1}
  \sigma_i^2 =  \int \left(x_i - \bar{x}_i \right)^2
  \pr{\bmath{x}}\mathrm{d}\bmath{x} 
  \approx \frac{1}{M}\sum_k \left(x_i^k - \bar{x}_i \right)^2,
\end{equation}
where \(k\) labels a sample in a chain of \(M\) samples and the
integral extends over the entire \(\bmath{x}\)-space. For our second
estimate we take the expression for the variance and integrate by
parts
\begin{eqnarray}
  \label{eq:varint}
  \lefteqn{\sigma_i^2 =
  \int  \left(x_i - \bar{x}_i \right)^2
  \pr{\bmath{x}}\mathrm{d}\bmath{x}}\nonumber \\
&= &
\frac{1}{3}
  \left|
    \left(
      x_i - \bar{x}_i 
    \right)^3\pr{x_i}
  \right|_{-\infty}^\infty \nonumber\\
  &&-\frac{1}{3}\int\left(x_i - \bar{x}_i \right)^3
  \frac{\upartial \pr{\bmath{x}}}{\upartial x_i} \mathrm{d}\bmath{x},
\end{eqnarray}
the first term of which will vanish if the marginalized distribution
\(\pr{x_i}\) drops off faster than \(x^{3}_i\) as \(x_i\) tends to
\(\pm\infty\). Using (\ref{eq:psi}) we rewrite this expression as
\begin{equation}
  \label{eq:varint2}
  \sigma_i^2 =\frac{1}{3}\int_{-\infty}^\infty \left(x_i - \bar{x}_i \right)^3
  \frac{\upartial \psi\left(\bmath{x}\right)}
  {\upartial x_i}\pr{\bmath{x}} \mathrm{d}\bmath{x}.
\end{equation}
We compute (\ref{eq:varint2}) from the samples in our chain by
\begin{equation}
  \label{eq:var2}
  \sigma^2_i \approx \frac{1}{M}\frac{1}{3} \sum_k 
   \left(x^k_i - \bar{x} \right)^3 \frac{\upartial \psi}{\upartial x_i}
   \bigg{\vert}_{x_i^k}.
\end{equation}
To test for convergence we compute the ratio \(R_i\) of
(\ref{eq:var1}) and (\ref{eq:var2}), and we believe the chain has
converged when all the \(R_i\) are close to unity.

We have tested how this criterion compares to the widely used
Gelman-Rubin statistic \citep{gelman1992} and have found that Hanson's
method tends to be, if anything, slightly pessimistic. We find that
values of \(R\) in the range \(0.8\) to \(1.2\) represent good
convergence and values in the range \(0.6\) to \(1.4\) are
acceptable. The Gelman-Rubin method requires multiple chains to be
generated and compares inter-chain with intra-chain statistics, whereas
Hanson's test uses a single chain and compares two different
intra-chain statistics. We use the Hanson test as it is very easy to
compute, scales well with problem size and requires that we only
generate one chain. We plan to explore other intra-chain convergence
diagnostics such as that proposed by \cite{Dunkley2005}.
\section{Hamiltonian Monte Carlo and power
  spectrum estimation}
\label{sec:hmcpsest} 
We use HMC to draw samples simultaneously from the joint density
(\ref{eq:joint}). Our potential is defined by \(\psi\left(\bmath{a},
  \{C_{\ell}\}\right) = - \log \pr{\bmath{a}, \{C_{\ell}\}|\bmath{d}}\) such
that
\begin{eqnarray}
\label{eq:potential}
\lefteqn{\psi\left(\bmath{a}, \{C_{\ell}\}\right) = 
\frac{1}{2}\left(\bmath{d}- \mat{Y}\mat{B}\bmath{a}\right)^{\rmn{T}}
\mat{N}^{-1}
\left(\bmath{d}- \mat{Y}\mat{B}\bmath{a}\right)} \nonumber 
\\&+&
\sum_\ell \left(\ell+\frac{1}{2}\right)
\left(\ln C_\ell+ \frac{\sigma_\ell}{C_\ell}\right) + \rmn{const}
\end{eqnarray}
and the gradient of the potential can be computed exactly by
\begin{equation}
  \label{eq:gradpota}
  \frac{\upartial\psi\left(\bmath{a}, \{C_{\ell}\}\right)}{\upartial \bmath{a}}
  = -\mat{B}\mat{Y}^{\rmn{T}} \mat{N}^{-1} \left(\bmath{d}- \mat{Y}\mat{B}\bmath{a}\right)
  +\mat{C}^{-1}\bmath{a}
\end{equation}
\begin{equation}
  \label{eq:gradpotc}
  \frac{\upartial\psi\left(\bmath{a}, \{C_{\ell}\}\right)}{\upartial C_{\ell}}
  = \left(\ell+\frac{1}{2}\right)\frac{1}{C_\ell}\left(1-\frac{\sigma_\ell}{C_\ell}\right).
\end{equation}

The positivity requirement on the power spectrum \(C_\ell\) can result in a high
rejection rate and we have found it advantageous to reparametrize the
problem in terms of the logarithm of the \(C_\ell\)s. For this
reparametrization it is easy to calculate the corresponding potential
and its derivatives. To enforce a flat prior on each \(C_\ell\) we
must apply an exponential prior on \(\log C_\ell\).

To generate a new sample requires us to evaluate the gradient at each
point along the leapfrog trajectory and to evaluate the value of the
potential once at the end of the trajectory. Therefore, if we take
\(n\) leapfrog steps, we must perform \(2n+1\) spherical harmonic
transforms, although we can reuse the gradient at the end of one
trajectory for the first step of the next.

We split the sampling process into a burn in phase, in which we
attempt to lose any dependence on our starting point, and a sampling
phase where we store the samples from the chain and we believe these
samples are drawn from the target density. During burn in we are
permitted to adjust the parameters of the sampler, for example to tune
the acceptance rate. Once burn in is complete we must fix the
parameters of the sampler in order that our samples come from the
desired distribution.

A good starting point can significantly reduce the time required for
burn in. We have explored a number of possibilities for computing a
starting point for the signal \(\bmath{a}\) given some initial guess
for the power spectrum. One we have found particularly effective is to
draw a single signal sample, as for one step of the Gibbs sampler,
from the conditional distribution \(\pr{\bmath{a}|\bmath{d},
  \{C_{\ell}\}}\). This is a computationally expensive process and is
described fully in \citet{Wandelt2004, Eriksen2004}. The basic
procedure involves solving the following equation for \(\bmath{x}\),
the spherical harmonic coefficients of the mean field (Wiener
filtered) map,
\begin{equation}
  \label{eq:meanfield}
  \left(\mat{C}^{-1} + \mat{B}\mat{Y}^{\rmn{T}}\mat{N}^{-1}\mat{Y}\mat{B}\right) \bmath{x}
  =
  \mat{B}\mat{Y}^{\rmn{T}}\mat{N}^{-1} \bmath{d}
\end{equation}
and a fluctuation term \(\bmath{y}\) that corrects for the bias in
\(\bmath{x}\)
\begin{equation}
  \label{eq:fluctuation}
  \left(
    \mat{C}^{-1} + \mat{B}\mat{Y}^{\rmn{T}}\mat{N}^{-1}\mat{Y}\mat{B}
  \right) \bmath{y}
  = \mat{C}^{-1/2}\omega_0  + \mat{B}\mat{Y}^{\rmn{T}}\mat{N}^{-1/2} \omega_1,
\end{equation}
where \(\omega_{0}\) is a set of spherical harmonic coefficients and
\(\omega_1\) a map both containing Gaussian white noise of zero mean
and unit variance. The sum of \(\bmath{x}\) and \(\bmath{y}\) is our
starting sample \(\bmath{a}\). We solve for \(\bmath{x}+\bmath{y}\)
using a conjugate gradient algorithm (see, for example
\citet{Golub1996}). A preconditioner can be used to reduce the number
of iterations required for the convergence of the conjugate gradient
solver, however the construction of a preconditioner is itself a
complex procedure, and since we only perform this step once and the
accuracy of the result is of little consequence, we have not made use
of one in this work. Whether or not applying the conjugate gradient
algorithm without a preconditioner is feasible depends on the nature
of the data set under consideration.

HMC has a large number of adjustable parameters, notably the
masses. The distribution for the \(\bmath{a}\) parameters is Gaussian
and so we attempt to set the mass associated with each \(a_{\ell m}\)
such that they are inversely proportional to the variance of that
\(a_{\ell m}\). We justify this choice in the Appendix. The masses for
the \(\bmath{a}\) are estimated for a fixed power spectrum for which
the variance is computed by
\begin{equation}
  \label{eq:varalm}
  \rmn{var}\left(a_{\ell m}\right) = 
  \left( 
    C_{\ell}^{-1} + 
    B_{\ell}N^{-1}_{\ell m, \ell^\prime m^\prime}\delta_{\ell \ell^\prime} \delta_{m m^\prime}B_{\ell}
  \right)^{-1},
\end{equation}
where we use our initial estimate of the power spectrum as the value of
\(C_\ell\) and compute the diagonal elements of the inverse noise
covariance matrix in harmonic space using Monte Carlo simulations.

At high \(\ell\) and with good signal-to-noise the marginal
distributions for each \(C_\ell\) are close to Gaussian and we can
obtain masses from the standard expression for the variance (see, for
example \cite{Zaldarriaga1997})
\begin{equation}
  \label{eq:varcl}
  \rmn{var}\left(C_{\ell}\right) = 
  \frac{2\ell + 1}{2f_{\rmn{sky}}} \left(C_{\ell} + N_{\ell}/B^2_{\ell}\right)^2,
\end{equation}
where \(N_{\ell}\) is the power spectrum of the noise in the
data,\(B_{\ell}\) is the beam transfer function and \(f_{\rmn{sky}}\)
is the fraction of the sky observed. For low multipoles the
distributions are significantly skewed and in low signal-to-noise the
sharp cut off of the distribution at \(C_\ell=0\) has a similar
effect. In these cases we have found that setting the masses from the
variances is insufficient. Instead we tune these masses
empirically. We aim to set the mass for each parameter to as small a
value as possible while maintaining our target acceptance rate. We
sample the \(C_\ell\)s from simple approximate likelihood function and
gradually reduce the masses until the acceptance rate drops. This
gives masses that are sufficient for sampling the full problem
efficiently.

During burn in we can further tune the masses; the convergence
criterion for each parameter providing a good indication of whether or
not the mass associated with that parameter is set correctly.

We must randomise the length of each trajectory and have found that
drawing \(n\) from a uniform distribution between \(10\) and \(20\) is
appropriate. Therefore we typically require the application of \(\sim
30\) spherical harmonic transforms to generate a new proposed
sample. We then tune the step size \(\epsilon\) such that we obtain an
acceptance rate between 70 and 90 per cent. A higher acceptance rate
is used for HMC than other MCMC methods as the computational cost of a
rejection is so high.

Once sampling we store each \(\{C_{\ell}\}\) sample and the
realization power spectrum \(\{\sigma_{\ell}\}\) of each signal
sample. The \(\{\sigma_{\ell}\}\) can be used to form the
Blackwell-Rao estimator of the posterior distribution
\citep{Chu2005}. The posterior can be written
\begin{eqnarray}
  \label{eq:brpost}
  \pr{\{C_{\ell}\}|\bmath{d}}&=&
  \int \pr{\{C_{\ell}\}, \bmath{a}|\bmath{d}}\rmn{d}\bmath{a}\nonumber\\
  &=&\int \pr{\{C_{\ell}\}|\bmath{a}}\pr{\bmath{a}|\bmath{d}}\rmn{d}\bmath{a},
\end{eqnarray}
which for a Gaussian CMB can be written as
\begin{equation}
  \label{eq:brsigma}
  \pr{\{C_{\ell}\}|\bmath{d}} = 
  \int \pr{\{C_{\ell}\}|\{\sigma_{\ell}\}} 
  \pr{\{\sigma_{\ell}\}|\bmath{d}}\rmn{d}\{\sigma_{\ell}\},
\end{equation}
where
\begin{equation}
  \label{eq:brpclgivensigma}
  \pr{\{C_{\ell}\}|\{\sigma_{\ell}\}} = \prod_{\ell}\frac{1}{\sigma_{\ell}}
  \left(
    \frac{\sigma_{\ell}}{C_{\ell}}
  \right)^{\frac{2 \ell + 1}{2}}
  \exp 
  \left(
    - \frac{2 \ell + 1}{2} \frac{\sigma_{\ell}}{C_{\ell}}
  \right).
\end{equation}
We can therefore compute the posterior probability of a set of \(\{C_{\ell}\}\)
from \(M\) samples \(\{\sigma_{\ell}^i\}\) by
\begin{equation}
  \label{eq:brfromsamples}
  \pr{\{C_{\ell}\}|\bmath{d}}\approx \frac{1}{M}\sum_i \pr{\{C_{\ell}\}|\{\sigma_{\ell}^i\}}.
\end{equation}
It is also possible to construct the marginal distributions for any
\(C_\ell\) or subset of \(\{C_\ell\}\). For a single \(C_\ell\) the
marginal distribution can be approximated by
\begin{equation}
  \label{eq:brmarginal}
   \pr{C_{\ell}|\bmath{d}}\approx \frac{1}{M}\sum_i \pr{C_{\ell}|\sigma_{\ell}^i},
\end{equation}
where
\begin{equation}
  \label{eq:brmarginal2}
  \pr{C_{\ell}|\sigma_{\ell}} = \frac{1}{\sigma_{\ell}}
  \left(
    \frac{\sigma_{\ell}}{C_{\ell}}
  \right)^{\frac{2 \ell + 1}{2}}
  \exp 
  \left(
    - \frac{2 \ell + 1}{2} \frac{\sigma_{\ell}}{C_{\ell}}
  \right).
\end{equation}
Extremely large numbers of samples would be needed to make this
estimator accurate at high \(\ell\). However even with a relatively
small number of samples it forms a useful tool for the analysis of
large angular scales. It is worth noting that the expression
(\ref{eq:brfromsamples}), or its one-dimensional marginalized version
(\ref{eq:brmarginal}), do not depend on the \(\{C_\ell\}\)-samples,
but only on the realization power spectra \(\{\sigma_\ell\}\) of the
\(\bmath{a}\)-samples.

\section{analysis of low-resolution simulations}
\label{sec:lowres}
To compare the Hamiltonian and Gibbs samplers we applied them both to
a set of low--resolution simulations. We produced a map of the CMB
with a HEALPix\footnote{ http://healpix.jpl.nasa.gov}
\(N_{\rmn{side}}=32\) (\(12288\) pixels). Our CMB simulation is a
realization of a \(\Lambda\)CDM cosmology with the best fitting
parameters from the 5-year \textit{WMAP}
observations\footnote{http://lambda.gsfc.nasa.gov/product/map/dr3/parameters.cfm}
\citep{Spergel2007} and includes multipoles up to \(\ell=64\). We
smoothed the map with a \(3\)-degree Gaussian beam and added isotropic
noise  with an RMS amplitude of \(55 \mu K\) per pixel. We
chose the noise level so that we could explore how the sampler
behaved as a function of the signal-to-noise ratio.  We degraded the
\textit{WMAP} Kp2 mask such that any (large) pixel in our final mask
is excluded if any of the (small) subpixels in the original mask are
excluded. This has the effect of enlarging the Kp2 mask to remove
around \(30\) percent of the sky: a large contiguous area along the
Galactic plane and a number of small regions around the locations of
bright point sources.

\begin{figure*}
  \centering
  \includegraphics[width=\textwidth]{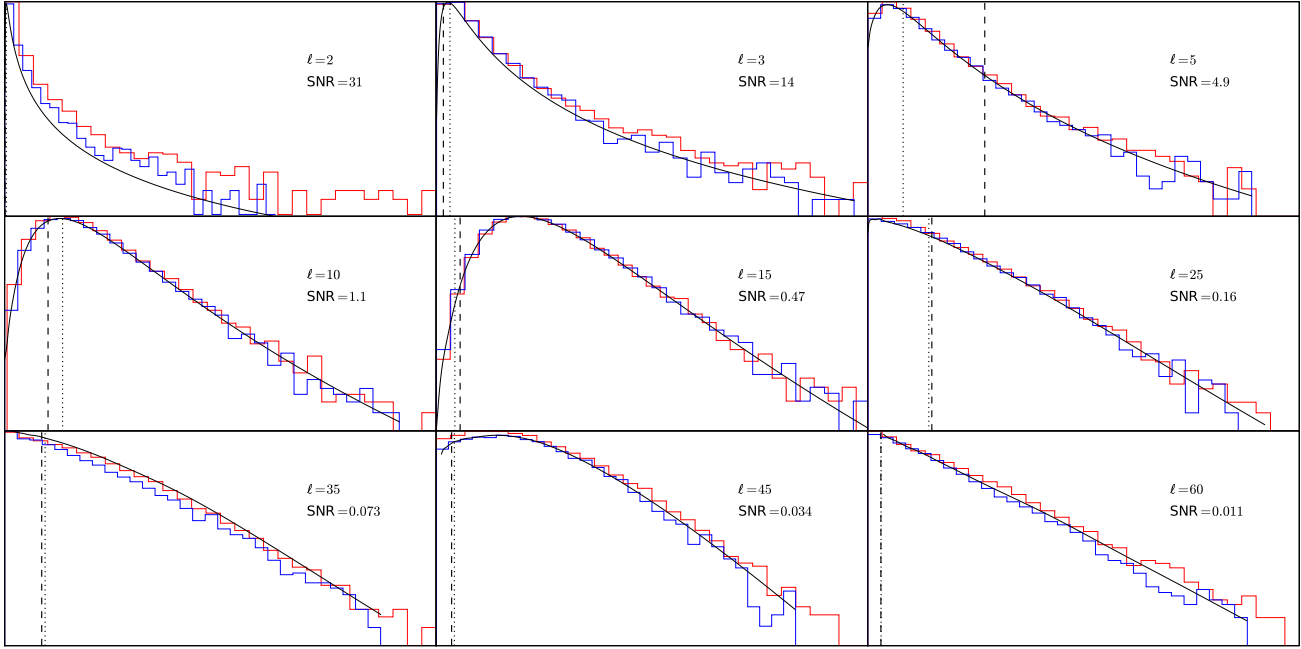}
  \caption{Marginal distributions of \(C_\ell\) samples for a
    selection of \(\ell\) as shown in the top right corner of each
    plot along with the signal to noise ratio for this multipole. The
    results for the Gibbs sampler are shown in red and the
    Hamiltonian sampler in blue. The plots show the logarithm of the
    number of samples falling in each bin. The dashed vertical line
    shows the theoretical value of the \(C_\ell\) used in creating the
    simulation whereas the dotted vertical line shows the value for
    the realization. The marginal distributions from the Blackwell-Rao
    estimator applied to the HMC samples are shown by the smooth black
    line.}
  \label{fig:lowresmarginal}
\end{figure*}

For each sampler we take 20000 burn in samples and then record the
next 50000 samples. A large number of samples helps to estimate
correlation lengths accurately; far fewer samples are required to
explore the distribution. The marginal distributions of a selection of
the \(C_\ell\) are shown in Fig. \ref{fig:lowresmarginal}. For most
\(\ell\) the data is too noisy to constrain the value of the
\(C_\ell\) however we do see good agreement between the results from
the Gibbs and Hamiltonian samplers. The HMC samples have also been used in
conjunction with the Blackwell-Rao estimator to generate a smooth
approximation to the marginal distributions. This estimator appears to
agree well with the histograms across this range of \(\ell\), but more
samples are likely to be needed if we were to calculate the joint
distribution of the \(\{C_\ell\}\).

\begin{figure*}
  \centering
  \includegraphics[width=\textwidth]{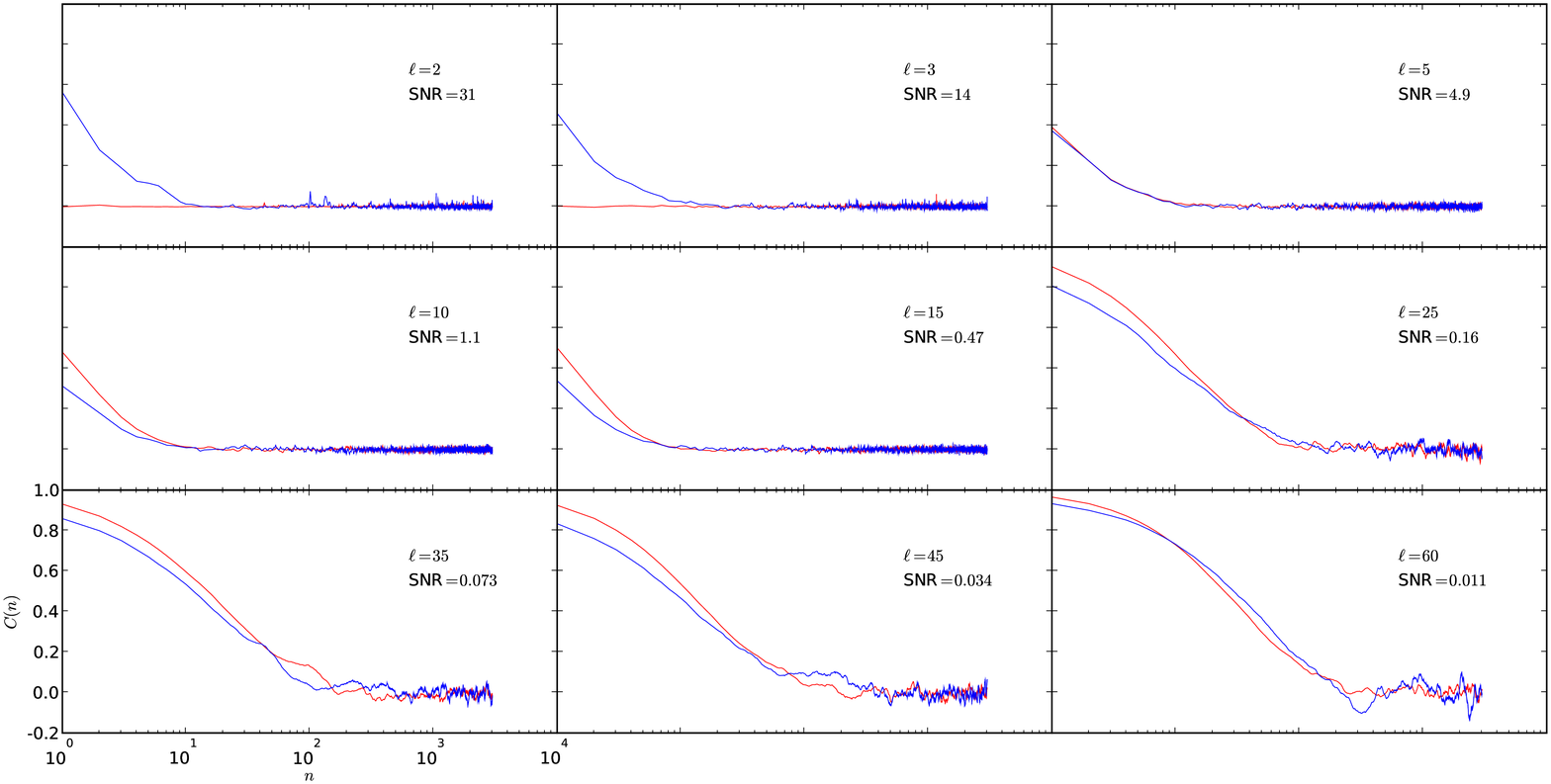}
  \caption{The auto-correlation functions (\ref{eq:autocor}) of
    \(C_\ell\) samples for a selection of \(\ell\) as shown in the top
    right corner of each plot along with the signal to noise ratio at
    this multipole. The results for the Gibbs sampler are plotted in
    red and Hamiltonian sampler in blue. All the plots use the same
    scale as shown in the bottom left plot.}
  \label{fig:lowrescorcoeff}
\end{figure*}

\begin{figure}
  \centering
  \includegraphics[width=0.5\textwidth]{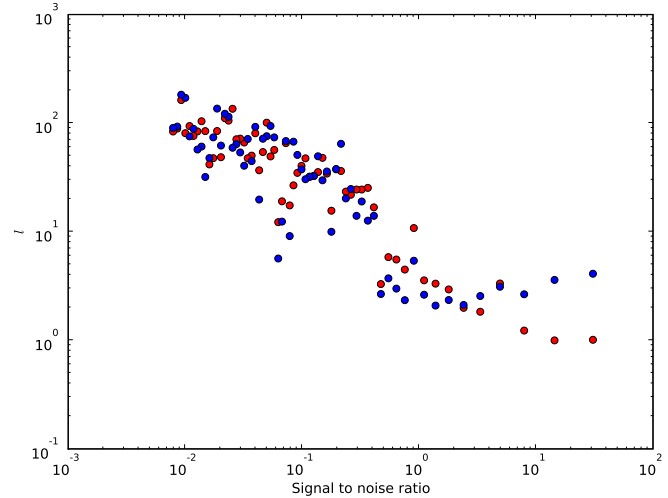}
  \caption{The correlation length (\ref{eq:corlen}) as a function of
    the signal-to-noise ratio of each \(C_{\ell} \) parameter. The
    red points show the results from the Gibbs sampler the blue
    points those from the Hamiltonian sampler. }
  \label{fig:lowrescorlen}
\end{figure}

In order to characterise the performance and efficiency of the
samplers we considered the correlation of the \(\{C_\ell\}\)
samples. Assuming that the \(C_\ell\)s are independent we can examine
the auto-correlation function,
\begin{equation}
  \label{eq:autocor}
  C(n) = 
  \left\langle 
    \frac{C_\ell^i - \langle C_\ell \rangle}{\sqrt{\mathrm{Var}(C_\ell)}}
    \frac{C_\ell^{i+n} - \langle C_\ell \rangle}{\sqrt{ \mathrm{Var}(C_\ell)}}
  \right\rangle.
\end{equation}
We show the auto-correlation function for a selection of multipoles in
Fig. \ref{fig:lowrescorcoeff}. As the signal-to-noise ratio for a
single \(\ell\), defined as the ratio of the signal and noise power
spectra at that \(\ell\), decreases with increasing \(\ell\) the
samples become more highly correlated; it takes more steps of the
samplers to generate independent samples. This feature is a well known
limitation of the Gibbs sampler caused by the fact that drawing the
power spectrum from the conditional distribution \(\pr{\{C_{\ell}\}|
  \mathbf{d}, \mathbf{a}}\) is limited to the size of the cosmic
variance while the joint distribution may be much wider. Similar
behaviour is observed with the Hamiltonian sampler although the cause
is now related to the difficulty in sampling the highly skewed
distributions that occur when the signal-to-noise ratio is low.
The correlation length for each parameter can be estimated using
\begin{equation}
  \label{eq:corlen}
  l = 1 + 2\sum_{n=1}^{n_{\mathrm{max}}}C(n),
\end{equation}
where we truncate the summation at some maximum lag
\(n_{\mathrm{max}}\) at which the auto-correlation function becomes
noisy. Fig. \ref{fig:lowrescorlen} shows how the measured correlation
lengths for the power spectrum parameters from the Gibbs and
Hamiltonian samplers depend on the signal-to-noise ratio for each
parameter, again estimated assuming the parameters are independent. We
see that in the high signal-to-noise regime the Gibbs sampler performs
exceptionally well whereas the Hamiltonian sampler produces samples
with typical correlation lengths of around four steps. Once the data
becomes noise dominated the picture is less clear with the Hamiltonian
sampler generally performing marginally better than the Gibbs
sampler. As the signal-to-noise ratio drops below about 0.01 both
samplers perform poorly.

It is worth noting that Hamiltonian sampler requires around an order
of magnitude fewer spherical harmonic transforms (the computationally
intensive step in the process) per sample than a Gibbs sampler that
uses no preconditioner and around a factor of 3-4 fewer transforms
than is reported for Gibbs samplers with carefully tuned
preconditioners \citep{Eriksen2004}. Furthermore we have found that
the correlation lengths of the Hamiltonian sampler strongly depend on
the masses one uses, offering the opportunity for significant
improvements given a more sophisticated prescription for setting the
masses.
\section{analysis of simulated \textit{WMAP} data}
\label{sec:wmap}
We produce a CMB simulation as for Section \ref{sec:lowres} but with
\(N_{\rmn{side}}=512\) (\(\sim 3\times 10^6\) pixels) and including
multipoles up to \(\ell=512\). The map was then smoothed with a \(13\)-arcmin Gaussian beam, which is similar in size to the beam of the
\textit{WMAP} \textit{W}-band. We then added anisotropic uncorrelated
noise by making use of the
published\footnote{http://lambda.gsfc.nasa.gov} \(N_{\rmn{obs}}\) and noise
variance for the 5-year \textit{WMAP} combined \textit{W} band
map. The map was cut with the Kp2 mask which excludes \(15.3\%\) of
the sky. We included multipoles up to \(\ell_{\rmn{max}} = 512\) in
our analysis. This gives us a total of around \(2\times 10^{5}\)
parameters in our sampling space.

To generate a good signal starting point, using a single Gibbs sample,
required \(\sim 800\) iterations (\(20\) minutes on the hardware
described below) of the conjugate gradient to solve
(\ref{eq:meanfield}) and (\ref{eq:fluctuation}) such that the rms
residual was less than \(10^{-6}\).

For these simulations we made a total of \(5000\) burn in samples and
recorded \(10000\) samples from the post burn-in phase. It takes
\(\sim 20\) seconds to generate a single sample using two dual core
Intel Xeon 5150 processors and the MPI parallelised HEALPix spherical
harmonic transforms, resulting in a total processing time of around
\(80\) hours.

\begin{figure}
  \centering
  \includegraphics[width=0.5\textwidth]{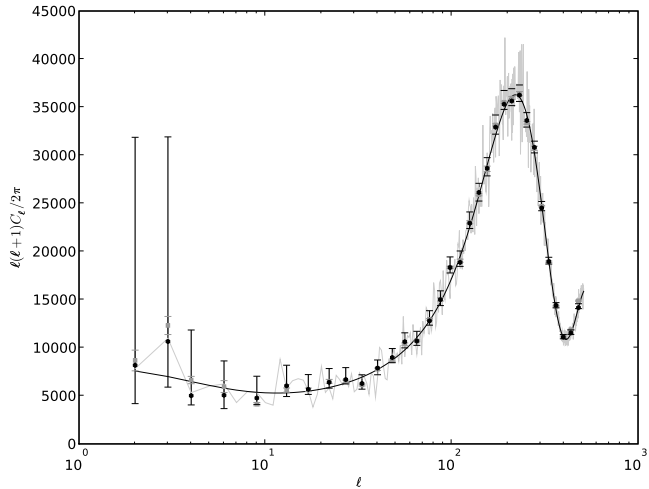}
  \caption{Binned power spectrum and \(68\) percent confidence
    intervals as compared to the results of an application of the
    MASTER method to the same simulated \textit{WMAP} data. The black
    solid line shows the power spectrum from which the simulation was
    generated while the grey shows the power spectrum of the
    realization. The grey squares and error bars show the MASTER
    results. The black circles and error bars show the peak and 68 per
    cent confidence intervals found from samples generated with the
    HMC sampler.}
  \label{fig:binnedps}
\end{figure}

\begin{figure}
  \centering
    \includegraphics[width=0.5\textwidth]{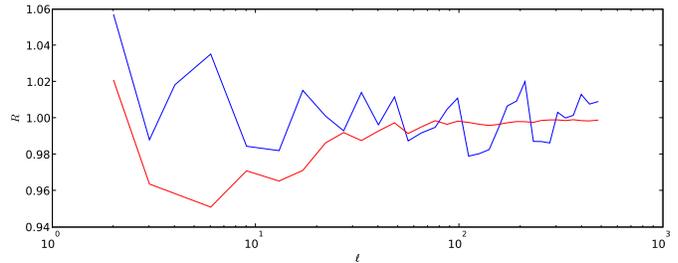}
    \caption{A summary of the convergence statistics of the \(10000\)
      samples used to produce the power spectrum in
      Fig. \ref{fig:binnedps}. Although convergence is judged from the
      \(R\) for every parameter we show here only the average \(R\)
      for in each bin for the \(C_{\ell}\) (blue line) and
      \(\bmath{a}\) (red line). \(R\) . }
  \label{fig:binnedR}
\end{figure}

For comparison we applied the MASTER method \citep{Hivon2002} to the
same data set. Our peak likelihood \(C_{\ell}\) sample and 68 per cent
confidence intervals, binned with the \textit{WMAP} team's scheme, are
shown alongside the results of the MASTER method in
Fig. \ref{fig:binnedps}. For most of the range of angular scales the
two estimates and their errors agree well. On the largest angular
scales the MASTER estimate tends to underestimate the uncertainities
and the symmetric errors are far from representative of the posterior.
In Fig. \ref{fig:binnedR} we show a summary of the convergence
statistics, using Hanson's diagnostic, see Section \ref{sec:converge},
demonstrating that we have fully explored the distribution across the
entire range in \(\ell\). For all multipoles the \(R\) value is within
the range \(0.9\) to \(1.1\).
\section{Conclusions}
\label{sec:conc}
We have introduced the HMC sampler for CMB power spectrum estimation
and demonstrated its performance both on low-resolution simulations
and simulations of 5-year \textit{WMAP} data. We find that the
Hamiltonian sampler has similar or shorter correlation lengths when
compared to the Gibbs sampler except in the regions of the highest
signal-to-noise. Bearing in mind the reduced computational cost and
greater flexibility of the Hamiltonian sampler we believe it is an
attractive method for performing the analysis.

For high-resolution data sets of size (\(N_{\rmn{side}}=512\),
\(\ell_{\rmn{max}}=512\)) we can generate a sample in \(\sim
20\) seconds on a high-end desktop. This is a significant gain over the
reported performance of Gibbs samplers.

HMC requires that we are able to compute the logarithm of the target
density and its gradients. Even if exact gradients are not available we
can generate approximate trajectories and these will still result in
samples drawn from the required distribution. The generality of the
approach removes the requirement for strictly Gaussian signal and
noise and therefore promises to be an interesting method for tackling
a wide range of related problems.

We are currently testing the performance of the method on
high-resolution \textit{Planck} simulations and working on extending
the method to include polarization. We also intend to apply the
technique to the \textit{WMAP} data.
\section{Acknowledgements}
\label{sec:acknowledgements}
We thank Morgan French for his contributions to the early development of our
sampler. JFT acknowledges a STFC (formerly PPARC) studentship. MAJA is
a member of the Cambridge Planck Analysis Centre, supported by STFC
grant ST/F005245/1 .This work was conducted in cooperation with
SGI/Intel utilising the Altix 3700 supercomputer at DAMTP Cambridge
supported by HEFCE and STFC. We acknowledge the use of the Legacy
Archive for Microwave Background Data Analysis (LAMBDA). Support for
LAMBDA is provided by the NASA Office of Space Science. Some of the
results in this paper have been derived using the HEALPix
\citep{Gorski2005} package.
\appendix
\section{masses for Hamiltonian Monte Carlo}
\label{sec:masshmc}
Hamiltonian Monte Carlo can be extremely sensitive to the choice of
masses. When sampling from an approximately isotropic distribution
this does not affect the performance significantly but when the
marginal distributions of different parameters show considerable
variation in width the masses must be set correctly to sample
efficiently. 

\citet{Hanson2001} suggests that one should set the mass associated
with each parameter to be approximately equal to the variance of that
parameter in the target density. This is an attempt to circularise the
trajectories in the \(\{x, p\}\) space. We take an alternative
approach, where the mass for a parameter is inversely proportional to
the width of the distribution, as suggested in \citet{Neal1996}. In
order to justify this approach we have generalised the framework in
\citet{Neal1993} to describe the application of the leapfrog
method. 

Consider the problem of sampling from an \(n\)-dimensional
Gaussian distribution in \(\bmath{x}\) with covariance matrix
\(\mat{C}\).  Our Hamiltonian is quadratic in \(\bmath{x}\) and
\(\bmath{p}\)
\begin{equation}
\label{eq:gauss:ham}
H = \frac{\bmath{p}^{\rmn{T}}\mat{M}^{-1}\bmath{p}}{2}+\frac{\bmath{x}^{\rmn{T}}\mat{C}^{-1}\bmath{x}}{2},
\end{equation}
where \(\mat{M}\) is a \(n\times n\) mass matrix, and the trajectory
will be determined by Hamilton's equations
\begin{equation}
  \label{eq:gausstrajx}
  \frac{\rmn{d}\bmath{x}}{\rmn{d} t} = \nabla_{\bmath{p}} H = \mat{M}^{-1} \bmath{p}
\end{equation}
\begin{equation}
  \label{eq:gausstrajp}
  \frac{\rmn{d} \bmath{p}}{\rmn{d} t} = -\nabla_{\bmath{x}} H = - \mat{C}^{-1}\bmath{x}.
\end{equation}
We integrate the equation of motion with the leapfrog method
\begin{equation}
  \label{eq:lfgauss1}
  \bmath{p}\left(t + \epsilon/2\right)=
  \bmath{p}\left(t\right) - 
  \frac{\epsilon}{2}\mat{C}^{-1}\bmath{x}\left(t\right)
\end{equation}
\begin{equation}
  \label{eq:lfgauss2}
  \bmath{x}\left(t + \epsilon\right)=
  \bmath{x}\left(t\right) + 
  \epsilon\mat{M}^{-1}\bmath{p}\left(t+\epsilon/2\right)
\end{equation}
\begin{equation}
  \label{eq:lfgauss3}
  \bmath{p}\left(t + \epsilon\right)=
  \bmath{p}\left(t+\epsilon/2\right) - 
  \frac{\epsilon}{2}\mat{C}^{-1}\bmath{x}\left(t+\epsilon\right).
\end{equation}
A single application of the leapfrog method can be written in the form
\begin{equation}
  \label{eq:lfgauss4}
  \bmath{x}\left(t+\epsilon\right) = 
  \left(\mat{I} - \frac{\epsilon^2}{2}\mat{M}^{-1}\mat{C}^{-1}\right)
  \bmath{x}\left(t\right) +
  \epsilon \mat{M}^{-1} \bmath{p}\left(t\right)
\end{equation}
\begin{eqnarray}
  \label{eq:lfgauss5}
  \bmath{p}\left(t+\epsilon\right)=&
  -\epsilon \mat{C}^{-1}
  \left(
    \mat{I} - \frac{\epsilon^2}{4}\mat{M}^{-1}\mat{C}^{-1}
  \right) \bmath{x}\left(t\right) + \nonumber\\
& +\left(\mat{I} - \frac{\epsilon^2}{2}\mat{C}^{-1}\mat{M}^{-1}\right)
  \bmath{p}\left(t\right), &
\end{eqnarray}
where \(\mat{I}\) is the identity matrix. We can rewrite this in a
matrix form
\begin{equation}
  \label{eq:lfgaussmat1}
  \left[\begin{array}{c}
      \bmath{x}\left(t+\epsilon \right)\\
      \bmath{p}\left(t+\epsilon \right)
    \end{array}
  \right]
  = \mat{T}
  \left[\begin{array}{c}
      \bmath{x}\left(t \right)\\
      \bmath{p}\left(t \right)
    \end{array}
  \right],
\end{equation}
where 
\begin{equation}
  \label{eq:lfgaussmat2}
  \mat{T} = \left[
    \begin{array}{cc}
      \left(\mat{I} - \frac{\epsilon^2}{2}\mat{M}^{-1}\mat{C}^{-1}\right) &
      \epsilon \mat{M}^{-1} \\
      -\epsilon \mat{C}^{-1}
      \left(
        \mat{I} - \frac{\epsilon^2}{4}\mat{M}^{-1}\mat{C}^{-1}
      \right) &
       \left(\mat{I} - \frac{\epsilon^2}{2}\mat{C}^{-1}\mat{M}^{-1}\right)
    \end{array}
    \right].
\end{equation}

If the method is to be stable under the repeated application of
\(\mat{T}\) then we require its eigenvalues to have unit
modulus. The eigenvalues \(\lambda\) are found from the characteristic
equation
\begin{equation}
\label{eq:characteristiceqn}
\det \left[
\mat{I}\lambda^2 - 
2\lambda \left(
\mat{I}- 	\frac{\epsilon^2}{2}\mat{M}^{-1}\mat{C}^{-1}
\right)
+\mat{I}
\right] = 0.
\end{equation}

To explore the space rapidly we wish to find the largest \(\epsilon\)
compatible with the condition for stability. Any dependence of
(\ref{eq:characteristiceqn}) on \(\mat{C}\) implies no single value
for \(\epsilon\) will meet the requirement for every eigenvalue to
have unit modulus (unless both \(\mat{C}\) and \(\mat{M}\) are
proportional to the identity matrix). The maximum value for
\(\epsilon\) should therefore be controlled by the width of the
distribution for a small subset of parameters.

By setting \(\mat{M} = \mat{C}^{-1}\) we remove the dependence of
\(\epsilon\) on the size of the distribution. In this situation the
characteristic equation reduces to
\begin{equation}
\label{eq:characteristiceqnsimple}
\left[
\lambda^2 - 
2\lambda \left(
1- 	\frac{\epsilon^2}{2}
\right)
+1\right]^n
 = 0
\end{equation}
and the stability criterion is met by \(\epsilon \le 2\).

If the dimensionally of the problem is such that it is impractical to
perform the required matrix inversion and decomposition of \(\mat{M}\)
(to compute the Hamiltonian and to draw new values for the momentum
variables respectively) then simple approximations must be
employed. Typically one might construct a diagonal mass matrix with
the mass associated with each parameter inversely proportional to
the variance of that parameter.

If the distribution to be sampled from is not Gaussian it seems
reasonable to use some appropriate measure of the width of the
distribution (i.e. the curvature at the peak \citep{Neal1996}) to set
the masses.

\label{lastpage}
\end{document}